\documentstyle[aps,prb,epsfig,preprint]{revtex}

\begin{document}
\draft
\tightenlines

\title{Theory of valence  transitions \\ 
                 in ytterbium-based compounds}

\author{V. Zlati\'c$^{1,3}$, and J. K. Freericks$^{2,3}$\\
$^1$ Institute of Physics, 10000 Zagreb, Croatia\\
$^2$Department of Physics, Georgetown University, Washington, DC 20057, USA\\
$^3$ Isaac Newton Institute, Cambridge CB3 0EH, UK}
\maketitle
\begin{abstract}
The anomalous behavior of YbInCu$_4$ and similar compounds is modeled 
by the exact solution of the spin one-half Falicov-Kimball model 
in infinite dimensions. 
The valence-fluctuating transition is related to a metal-insulator 
transition caused by the Falicov-Kimball interaction, and triggered by 
the change in the f-occupancy. 
\end{abstract}

\section{Introduction}
The intermetallic compounds of the YbInCu$_4$ family exhibit an 
isostructural transition from high-temperature 
state with trivalent Yb ions in the $4f^{13}$ configuration to 
the low-temperature  mixed-valent state with Yb ions fluctuating 
between $4f^{13}$ and $4f^{14}$ configurations~\cite{felner.86}. 
The transition is particularly abrupt in high-quality 
stoichiometric YbInCu$_4$ samples~\cite{sarrao.96}  
with a transition temperature equal to $T_v=42$~K at ambient pressure; 
the susceptibility and the resistivity drop at $T_v$ by more 
than one order of magnitude in cooling, 
while the volume expansion is small, $\Delta V/V\simeq 0.05$. 
The valence change  inferred from $\Delta V/V$ by using 
the usual ionic radii of Yb$^{3+}$ and Yb$^{2+}$ 
is about $\Delta n_f\simeq 0.1$, which is consistent 
with the valence measurements by the $L_{III}$-edge 
absorption~\cite{felner.86,cornelius.97}.  
The critical temperature depends strongly on external pressure, 
magnetic field, and alloying~\cite{immer.97,sarrao.98}. 
A recent review of the experimental data is given in Ref.~\cite{sarrao.99} 
and here we just recall the main points which motivate 
our choice of model. 

The integer-valent phase ($T\geq T_v$) is characterized by a 
Curie-Weiss susceptibility~\cite{felner.86,sarrao.98}
with very small Curie-Weiss temperature $\Theta\ll T_v$. 
The Curie constant corresponds to the free moment of one 
magnetic f-hole in a $J=7/2$ spin-orbit state with
$\mu_{eff}=4.53 \mu_B$. 
The electrical resistance is large and has a small positive slope;
it remains almost unchanged in magnetic fields up to 30 T~\cite{immer.97}. 
In some systems, like Yb$_{1-x}$Y$_x$InCu$_4$, 
the magnetoresistance is slightly negative, 
while in YbInCu$_4$ (or YbIn$_{1-x}$Ag$_x$Cu$_4$ for $x=0.15$)  
it is slightly positive~\cite{immer.97}. 
The Hall constant is large and negative, indicating a small number of 
carriers~\cite{cornelius.97,figueroa.98}.
The thermoelectric power has a rather small slope which one finds 
in a semiconductor with a nearly symmetric density of states~\cite{ocko.2000}. 
Recent data on the optical conductivity of YbInCu$_4$~\cite{garner.00}  
shows the absence of a Drude peak at high temperatures and 
a pronounced maximum of the optical spectral weight at about 1 eV.
The high-temperature ESR data for Gd$^{3+}$ embedded in YbInCu$_4$ resemble
those found in integer-valence semi-metallic or insulator 
hosts~\cite{altshuler95}. Thus, the high-temperature phase 
indicates the presence of a well defined local moment but 
gives no signature of the Kondo effect. The overall behavior of the 
high-temperature phase is closer to that of a semi-metal 
or paramagnetic small-gap semiconductor than to a Kondo metal.  

The mixed-valent phase ($T < T_v$) behaves like a Pauli paramagnet 
with moderately enhanced susceptibility and specific heat 
coefficient~\cite{sarrao.98}.
The electrical resistance and the Hall constant are one order of magnitude 
smaller than in the high-temperature phase~\cite{cornelius.97,figueroa.98}.
The thermoelectric power~\cite{ocko.2000} has a very large slope  
typical of a valence fluctuator with large asymmetry 
in the density of states.  
The susceptibility, the resistivity and the Hall constant do not show 
any temperature dependence below $T_v$, which is also typical 
of valence fluctuators. 
The optical conductivity shows a major change with respect to 
the high-temperature shape. The peak around 1 eV is reduced,  
the Drude peak becomes fully developed, and an additional 
structure in the mid-infrared range appears quite suddenly 
below $T_v$~\cite{garner.00}.
A large density of states at the chemical potential $\mu$ is indicated 
by the ESR data as well~\cite{rettori.97}.  
Thus, the  transition at $T_v$ seems to be from a paramagnetic 
semimetal to a valence fluctuator. 

In contrast to usual valence-fluctuators, which are quite 
insensitive to the  magnetic field, the  YbInCu$_4$ family 
of compounds also exhibit metamagnetic transitions when $T<T_v$. 
The Yb moment is fully restored at a critical field $H_c(T)$,
with a Zeeman energy $\mu_B H_c$  comparable to the thermal energy $k_BT_v$.
The metamagnetic transition defined by the magnetoresistance 
or the magnetization data~\cite{sarrao.99} 
gives an H-T phase boundary $H_c(T)=H_c^0\sqrt{1-(T/T_v)^2}$. 
The zero-temperature field $H_c^0$ is related to $T_v$ 
as $k_B T_v/\mu_B H_c^0=1.8$~\cite{sarrao.99}. 

To account for these features we need a model in which the non-magnetic, 
valence-fluctuating, metallic ground state can be 
destabilized by increasing temperature or magnetic field. 
Above the transition, we need a paramagnetic  
semiconductor with an average f-occupancy that is not changed much 
with respect to the ground state. 
The correct model for this system is a periodic Anderson model supplemented 
with a large Falicov-Kimball (FK) interaction term.  
The temperature or field induced transition suggests that 
one should place the narrow f-level just above the chemical potential $\mu$.
The hybridization keeps the f-count finite below the transition, 
while large f-f correlations allow only the fluctuations between 
zero- and one-hole (magnetic) configurations. 
The low-temperature phase is close to the valence fluctuating fixed 
point and shows no Kondo effect.  
However, because of the Falicov-Kimball term, there is a critical 
f-occupation at which there is a transition into the high-temperature 
state with a large gap in the d- and f-excitation spectrum.
The $n_f$ is driven to criticality either by temperature or magnetic field. 
In the high-temperature phase the hybridization can be neglected 
because the f-level width is already large due to thermal fluctuations, 
and quantum fluctuations are irrelevant. 
Unfortunately, the above model would be difficult to solve 
in a controlled way, and here we consider a simplified model  
in which the hybridization is neglected at all temperatures.  
This leads to a spin-degenerate Falicov-Kimball model which 
explains the collapse of the non-magnetic metallic phase 
at $T_v$ or $H_c$, and gives a good qualitative description 
of the high-temperature paramagnetic phase. 
However, the deficiency of the simplified model is that it yields 
a negligible f-count in the metallic phase and predicts a large 
change in the Yb valence at $T_v$ or $H_c$. 
It is clear that we can not obtain the valence fluctuating 
ground state and maintain the average f-occupancy 
below the transition without hybridization-induced quantum fluctuations.  
In what follows, we describe the model, explain the method of 
solution, and present results for static and dynamic correlation 
functions. 
 
\section{Calculations}

The Hamiltonian of the Falicov-Kimball model~\cite{falicov.69} 
consists of two types of electrons:  
conduction electrons (created or destroyed at site $i$
by $d_{i\sigma}^{\dagger}$ or $d_{i\sigma}$) and localized electrons 
(created or destroyed at site $i$ by $f_{i\sigma}^{\dagger}$ 
or $f_{i\sigma}$). The conduction electrons can hop 
between nearest-neighbor sites on the D-dimensional lattice, 
with a hopping matrix  $-t_{ij}=-t^*/2\sqrt{D}$; 
we choose a scaling of the hopping matrix that yields a nontrivial limit
in infinite-dimensions~\cite{metzner_vollhardt}.  
The $f$-electrons have a site energy $E_f$, 
and a chemical potential $\mu$ is employed to conserve the 
total number of electrons $n_{d\uparrow}+n_{d\downarrow}+
n_{f\uparrow}+n_{f\downarrow}=n_{tot}$.  
The Coulomb repulsion $U_{ff}$ between two $f$-electrons is infinite
and there is a Coulomb interaction $U$ between the $d$- and $f$-electrons
that occupy the same lattice site.  
An external magnetic field $h$ couples to 
localized electrons with a Land\'e g-factor.  
The resulting Hamiltonian is~\cite{brandt.89,freericks.98}
\begin{eqnarray}
H=\sum_{ij,\sigma }
(-t_{ij}-\mu \delta _{ij})d_{i\sigma }^{\dagger}d_{j\sigma }
+
\sum_{i,\sigma }(E_f-\mu )f_{i\sigma }^{\dagger }f_{i\sigma }\cr 
+
U\sum_{i,\sigma \sigma ^{\prime }}d_{i\sigma }^{\dagger }
d_{i\sigma}f_{i\sigma ^{\prime }}^{\dagger }f_{i\sigma ^{\prime}}
+
U_{ff}\sum_{i,\sigma }f_{i\uparrow }^{\dagger }
f_{i\uparrow}f_{i\downarrow }^{\dagger }f_{i\downarrow }\cr 
-
\mu _Bh\sum_{i,\sigma}\sigma (2d_{i\sigma }^{\dag }d_{i\sigma }
+gf_{i\sigma }^{\dag }f_{i\sigma}).  
                                      \label{eq:H_FK}  
\end{eqnarray}
The model can be solved in the infinite-dimensional limit 
by using the methods of Brandt-Mielsch~\cite{brandt.89}.
We consider the hypercubic lattice with Gaussian density 
of states $\rho(\epsilon)
=
\exp [-\epsilon^2/t^{*2}]/\sqrt{\pi}t^*$,
and take $t^*$ as the unit of energy ($t^*=1$). 
Our calculations are restricted to the homogeneous phase.

The local conduction-electron Green's function satisfies Dyson's equation
\begin{equation}
G^{\sigma}(z) =\int \frac{\rho(\epsilon)}
{z+\mu-\Sigma^{\sigma}(z)-\epsilon}d\epsilon,
                                            \label{eq: gloc}
\end{equation}
where $z$ is a complex variable and $\Sigma^{\sigma}$ 
is the local self energy which does not depend on 
momentum~\cite{metzner_vollhardt}. 
In infinite dimensions, $\Sigma^{\sigma}$ is defined 
by a sum of skeleton diagrams, which depend on the local 
d-propagator $G^{\sigma}$ but not on $t_{ij}$. 
The exact self-energy functional for the FK model 
is obtained by calculating the thermodynamic Green's 
function~\cite{kadanoff-baym} of an atomic system coupled 
to an external time-dependent field 
$\lambda^{\sigma}(\tau)$  
\begin{equation}
G^{\sigma}_{\rm atom}(\tau)
= -\frac{1}{\cal Z}
{\rm Tr}_{df} \left<
T_{\tau}
e^{-\beta H_{\rm atom}}
d_{\sigma}(\tau^{\prime}) d_{\sigma }^{\dagger}(\tau)
S(\lambda) \right> ,
\end{equation}
where the S-matrix for the $\lambda$-field is 
\begin{equation}
S(\lambda)
=
T_{\tau}
e^{-
\int_0^{\beta}d\tau\int_0^{\beta} d\tau^{\prime}\lambda(\tau,\tau^{\prime})
d_{\sigma }^{\dagger}(\tau)d_{\sigma}(\tau^{\prime})
} ,
\end{equation}
and $H_{\rm atom}$ is obtained from the Hamiltonian (\ref{eq:H_FK}) 
by removing the hopping and keeping just a single lattice site. 
The exact solution for $G_{n}^{\sigma}[\{\lambda_m\}]$ 
at Matsubara frequency $i\omega_n=i\pi T(2n+1)$ is given by, 
\begin{equation}
G_{n}^{\sigma}=
\frac{w_0}{[G^{\sigma}_{0n}]^{-1}}
+ 
\frac{w_1}{[G^{\sigma}_{0n}]^{-1}-U} ,
                          \label{G-atomic}
\end{equation}
where  $w_0$ and $w_1$ are the f-occupation numbers ($w_1=1-w_0$,
$w_0 = {\cal Z}_0/\cal Z$) 
and~\cite{brandt.89}
\begin{equation}
{\cal Z}_0(\lambda,\mu)
=
2e^{\beta\mu/2}\prod_n \frac{1}{(i\omega_n)G^{\uparrow}_{0n}}
2e^{\beta\mu/2}\prod_n \frac{1}{(i\omega_n)G^{\downarrow}_{0n}},
\end{equation}
with
\begin{equation}
{\cal Z}(\lambda,\mu)={\cal Z}_0(\lambda,\mu)
+
2e^{-\beta (E_f-\mu)}{\cal Z}_0(\lambda,\mu-U).
\end{equation}
The bare Green's function satisfies
\begin{equation}
G_{0n}^{\sigma}
=
\frac{1}{i\omega_n+\mu - \lambda^{\sigma}_n},
\end{equation}
with $\lambda_n$ the Fourier transform of the external time-dependent field.

The self-energy functional $\Sigma_{n}^{\sigma}[G_{n}^{\sigma}]$ 
can now be obtained~\cite{brandt.89} by using the Dyson equation  
for the atomic propagator, 
\begin{equation}
\Sigma_{n}^{\sigma}
=
[G_{0n}^{\sigma}]^{-1} 
-
[G_{n}^{\sigma}]^{-1} ,  
                         \label{Dyson}
\end{equation}
and eliminating $G_{0n}^{\sigma}[\{\lambda_m\}]$ from 
Eqs.~(\ref{G-atomic})  and (\ref{Dyson}).
The mapping onto the lattice is achieved by  
adjusting  $G_{0n}^{\sigma}$ in such a way that 
$G_{n}^{\sigma}[\{\lambda_m\}]$ satisfies the lattice Dyson 
equation (\ref{eq: gloc}). 

The numerical implementation of the above procedure is as follows: 
We start with an initial guess for the self energy $\;\Sigma^{\sigma}\;$   
and calculate the local propagator in (\ref{eq: gloc}). 
Using (\ref{Dyson}) we calculate the bare atomic propagator 
${G_{0n}^{\sigma}}$ and find ${\cal Z}_0$ and  $\cal Z$. 
Next we obtain $w_0$, $w_1$ and find $G_{n}^{\sigma}$ from (\ref{G-atomic}). 
Using ${G_{0n}^{\sigma}}$ and $G_{n}^{\sigma}$, we compute the 
atomic self energy  and iterate to the fixed point. 

The iterations on the  imaginary axis give static properties, like 
$n_f$,  the f-magnetization $m_f(h,t)$, and
the static spin and charge susceptibilities. 
Having found the f-electron filling $w_1$
at each temperature, we iterate Eqs.~(\ref{eq: gloc}) 
to (\ref{Dyson}) on the real axis and obtain the retarded dynamical properties, 
like the spectral function, the resistivity, 
the magnetoresistance, and the optical conductivity.  
At the fixed point, the spectral properties of the atom perturbed by 
$\lambda$-field coincide with the local spectral properties of the 
lattice. 

\section{Results and discussion}

We studied the model for a total electron filling of 1.5 and for 
several values of $E_f$ and $U$. 
The main results can be summarized in the following way.

\begin{figure}[htb]
\center{\epsfig{file=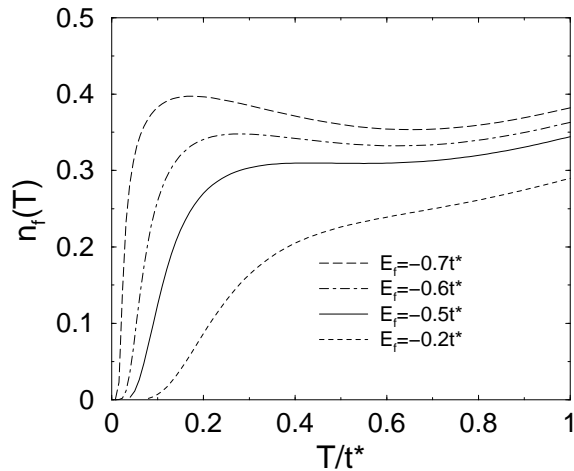,height=75mm,angle=-90}}
\caption{
Number of the $f$-holes plotted versus $T/t^*$ 
for $U/t^*=4$. The $E_f/t^*$ increases from 
top to bottom, and is given by -0.7, -0.6, -0.5, 
and -0.2, respectively.}
                                          \label{n_vs_T}
\end{figure}
The occupancy of the f-holes at high temperatures is large and there is 
a huge magnetic degeneracy. The f-holes are energetically unfavorable 
but are maintained because of their large magnetic entropy. 
In Fig.(\ref{n_vs_T}) we show $n_f$  as a function of temperature, 
plotted for $U=4t^*$, and $E_f/t^*$ from -0.2 to -0.7. 
Below a certain temperature, which depends on $U$ and $E_f$,
there is a rapid transition to the low-temperature phase. 
The transition becomes sharper and is pushed to lower temperatures 
as $E_f$ decreases at constant $U$. 
However, we restrict ourselves to continuous crossovers here, 
since the region with first-order transitions leads to 
numerical instabilities. 

The uniform f-spin susceptibility is obtained by calculating the spin-spin 
correlation function~\cite{brandt.89,freericks.98} and is given by  
$\chi(T)=C n_f(T)/T$, where $C=g_L^2\mu_B^2 J(J+1)/3k_B$ is the Curie 
constant. 
\begin{figure}[htb]
\center{\epsfig{file=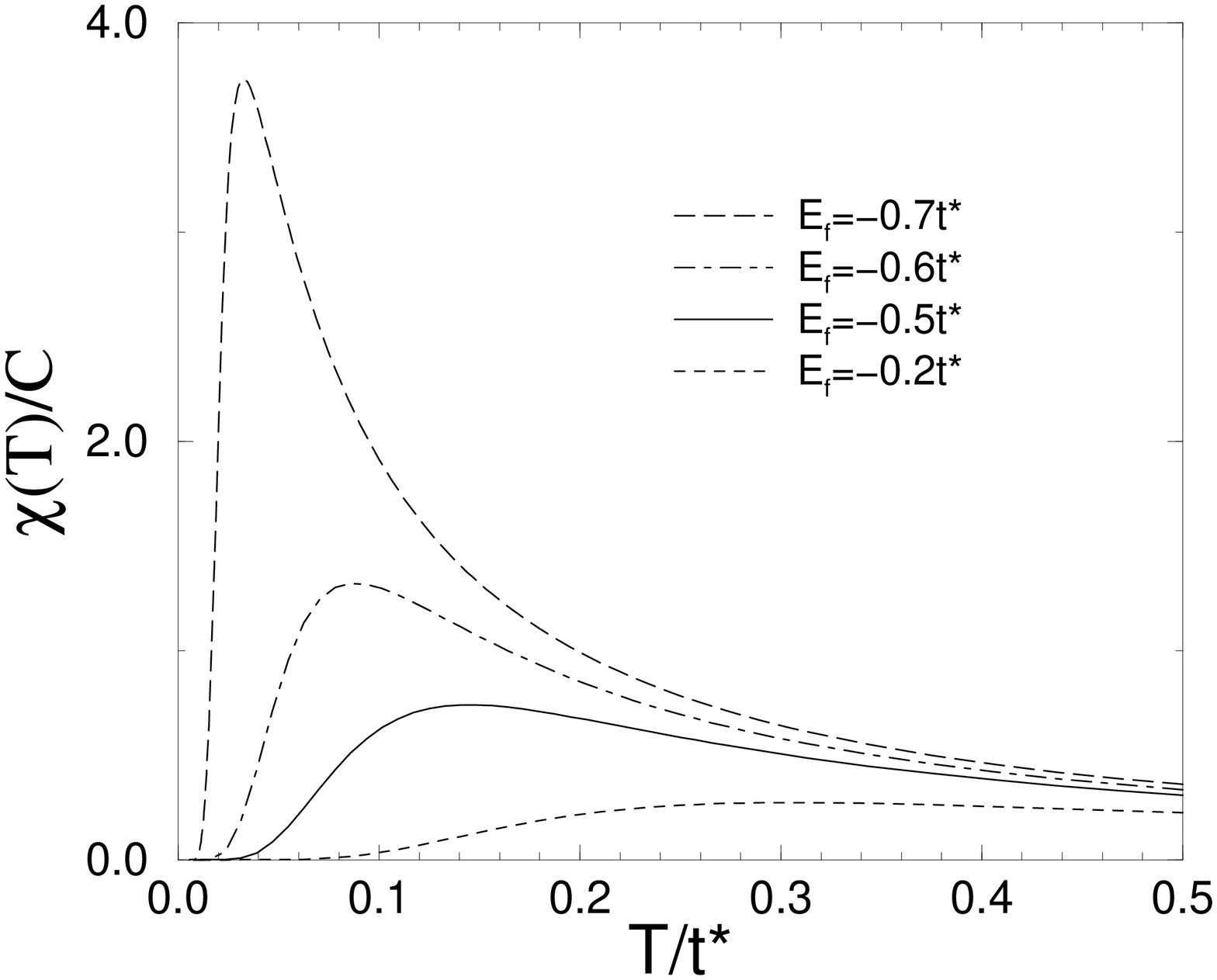,height=75mm,angle=-90}}
\caption{Uniform static magnetic susceptibility 
of the $f$-holes plotted versus $T/t^*$ for $U/t^*=4$. 
The values of $E_f/t^*$ are the same as in Fig(\ref{n_vs_T}).
The corresponding values of $T_v/t^*$ are 
estimated from the maximum of $\chi(T)$, 
and are given by 0.03, 0.08, 0.15, 0.35, respectively. 
The $T_v$ increases from top to bottom.}
                                        \label{chi_s_vs_T}
\end{figure}
The $\chi(T)/C$ is shown in Fig.~(\ref{chi_s_vs_T}) for $U/t^*=4$ 
and for $E_f$ as quoted in Fig.(\ref{n_vs_T}). 
The $T_v$ is obtained from the maximum of the $\chi(T)/C$ and  
the values corresponding to various parameters used in 
this paper are quoted in the caption of Fig.~(\ref{chi_s_vs_T}). 
The high-temperature susceptibility follows an approximate 
Curie-Weiss law, but the Curie-Weiss parameters depend 
on the fitting interval. 

\begin{figure}[htb]
\center{\epsfig{file=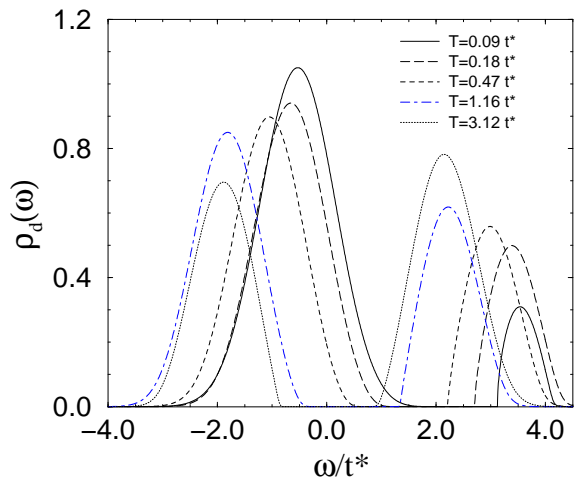,height=75mm,angle=-90}}
\caption{
Interacting density of states plotted versus 
$\omega/t^*$ for $U/t^*=4$, $E_f/t^*=-0.5$ ($T_v/t^*=0.14$), 
and for various temperatures, as indicated in the figure.}
                                      \label{dos_vs_omega}
\end{figure}

The interacting density of states $\rho_d(\omega)$ for the conduction 
electrons is shown in Fig.(\ref{dos_vs_omega}) for $U/t^*=4$ and 
$E_f/t^*=-0.5$, and for several temperatures.
(The energy is measured with respect to $\mu$.) 
The  high-temperature DOS has a gap of the order of $U$, 
and the chemical potential is located within the gap. 
Below the transition $n_f$ is small, the correlation effects 
are reduced, and $\rho_d(\omega)$ assumes a nearly non-interacting 
shape, with large $\rho_d(\mu)$  and halfwidth $W\simeq t^*$.

The transport properties of the high-T phase are dominated by the 
presence of the gap, which leads to a small dc conductivity 
with a weak temperature dependence. 
The transport properties of the paramagnetic 
phase are unrelated to the spin-disorder Kondo scattering 
(there is no spin-spin scattering in the FK model). 
Below the transition the conductivity increases and 
assumes large metallic values. 

\begin{figure}[htb]
\center{\epsfig{file=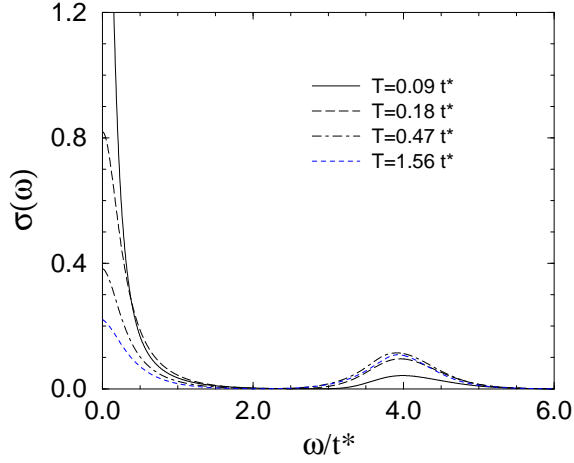,height=75mm,angle=-90}}
\caption{Optical conductivity 
plotted versus $\omega/t^*$ 
for various temperatures.
The $U$, $E_f$, and $T_v$, are the same as 
in Fig.(\ref{dos_vs_omega}).}
                                 \label{optical_vs_omega}
\end{figure}

The intraband optical conductivity $\sigma(\omega)$ is plotted in 
Fig.(\ref{optical_vs_omega}) 
as a function of frequency, for several temperatures. 
Above $T_v$, we observe a reduced Drude peak around $\omega=0$ 
and a pronounced high-frequency peak around $\omega\simeq U$. 
The shape of $\sigma(\omega)$ changes completely across $T_v$. 
Below $T_v$ the  Drude peak is fully developed and there 
is no high-energy (intraband) structure. 
However, if the renormalized f-level is close to $\mu$, 
the interband d-f transition could lead to an additional 
mid-infrared peak.
The ratio of the high-frequency peak in Fig(\ref{optical_vs_omega})
and the corresponding value of $T_v=0.15t^*$, is $U/T_v=26$. 
For the same value of U and $E_f=-0.7t^*$ ($T_v=0.03t^*$) 
we obtain $U/T_v=130$, while for $E_f=-0.75t^*$ ($T_v=0.02t^*$) 
we find $U/T_v\simeq 200$ (not shown). 
If we estimate the f-d correlation in YbInCu$_4$ from the 8000 $cm^{-1}$ 
peak in the optical conductivity data~\cite{garner.00},  
we obtain the experimental value $U\simeq 1$ eV. Together with 
$T_v=42$ K~\cite{sarrao.99} this gives the ratio $U/T_v\simeq 200$. 
If we take $U/t^*=4$ and adjust $E_f/t^*$ so as to bring 
the theoretical value of $T_v$ in 
agreement with the the thermodynamic and transport data on YbInCu$_4$, 
we get a high-frequency peak in $\sigma(\omega)$ at about 
8000 $cm^{-1}$, 6000 $cm^{-1}$, and 1500 $cm^{-1}$,   
for $E_f=-0.75t^*$, $E_f=-0.7t^*$, and $E_f=-0.5t^*$, respectively.

\begin{figure}[htb]
\center{\epsfig{file=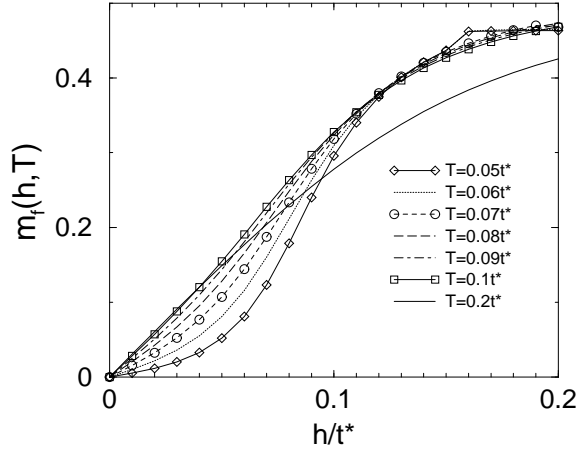,height=75mm,angle=-90}}
\caption{The f-electron magnetization 
$m_f$ is plotted as a function of $h/t^*$ for various temperatures.
The $U$, $E_f$, and $T_v$, are the same as  
in Fig.(\ref{dos_vs_omega}).}
                                        \label{m_vs_H}
\end{figure}
The f-electron magnetization $m_f(h)$ is plotted in Fig.(\ref{m_vs_H}) 
versus reduced magnetic field $h/t^*$, for several temperatures.  
Above the characteristic temperature $T_v^*\simeq T_v/2$,  
the $m_f(h)$ curves exhibit typical local moment behavior. 
Below $T_v^*$ we find a metamagnetic transition at a 
critical field $H_c$; the $m_f(h)$ is negligibly small below $H_c$ and
the local moment is fully restored above $H_c$. 
\begin{figure}[htb]
\center{\epsfig{file=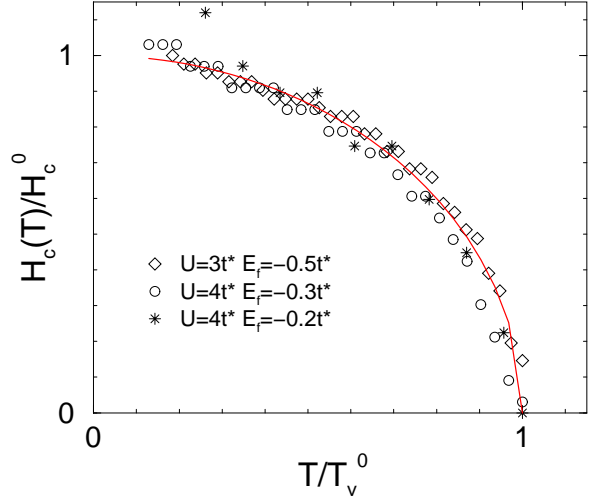,height=75mm,angle=-90}}
\caption{Normalized critical field is plotted 
as a function of reduced temperature $T/T_v^*$ 
for several values of $E_f/t^*$ and $U/t^*$. 
The full line represents $\sqrt{1-(T/T_v^*)^2}$ and $T_v^*=T_v/2$.
}
                                     \label{Hc_vs_T}
\end{figure}
Taking the inflection point of the $m_f(h)$ curves,  
calculated for several values of U and $E_f$, as an estimate
of $H_c(T)$ we obtain the phase boundary which is shown 
in Fig.(\ref{Hc_vs_T}), together with the expression 
$H_c(T)/H_c^0=\sqrt{1-(T/T_v^*)^2}$. 
Note, the $T_v^*$ values in Fig.(\ref{Hc_vs_T}) differ by more than 
an order of magnitude, while the ratio $k_B T_v^*/\mu_B H_c^0$ 
is only weakly parameter dependent. 

\begin{figure}[htb]
\center{\epsfig{file=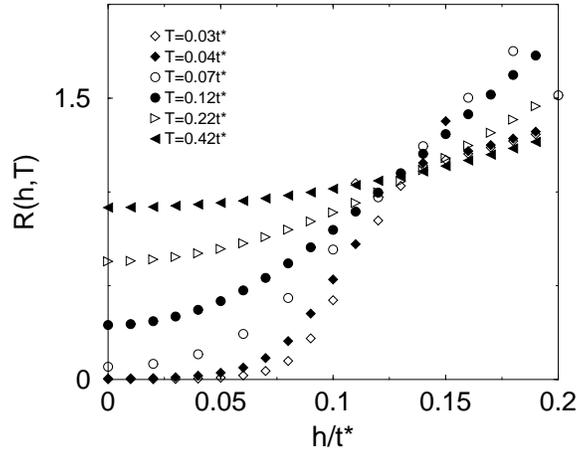,height=75mm,angle=-90}}
\caption{Field-dependent resistivity 
plotted versus $h/t^*$. The different symbols 
correspond to different temperatures, 
as indicated in the figure.
The $U$ and $E_f$ are the same as 
in Fig.(\ref{dos_vs_omega}).}
                                    \label{h_resistance_vs_h}
\end{figure}
The metamagnetic transition is also seen in the field-dependent 
electrical resistance $R(h,T)$ which is plotted in 
Fig.(\ref{h_resistance_vs_h}) as a function of $h/t^*$, 
for several temperatures. A substantial change in the $R(h,T)$ 
across $T_v^*$ or $H_c$ is clearly seen.
\section{Summary}
From the preceding discussion it is clear that Falicov-Kimball model 
captures the main features of the experimental data for YbInCu$_4$ 
and similar compounds. The temperature- and field-induced anomalies 
are related to a metal-insulator transition, 
which is caused by large FK interaction and triggered by 
the temperature- or the field-induced change in the f-occupancy. 
At high temperatures, we find a large gap in $\rho_d(\omega)$;
we expect a similar gap in the f-electron spectrum as well. 
At low temperatures, both gaps are closed, and the renormalized 
f-level renormalizes down to the chemical potential.

Our calculations describe doped Yb systems with broad transitions 
but appear to be less successful for those compounds which 
show a first-order transition. The numerical curves can be made sharper
(by adjusting the parameters) but they only become discontinuous 
in a narrow parameter range. 
The main difficulty with the FK model is that it predicts a substantial 
change in the f-occupancy across the transition and associates the 
loss of moment with the loss of f-holes. 
But in the real materials the loss of moment seems to be due to the 
valence fluctuations, rather than to the reduction of $n_f$. 
The description of the valence fluctuating ground state would 
require the hybridization and is beyond the scope of this work. 
The actual situation pertaining to Yb ions in the mixed-valence state might 
be quite complicated,  since one would have to consider 
an extremely asymmetric limit of the Anderson model, 
in which the ground state is not Kondo-like, there is no Kondo resonance, 
and there is no single universal energy scale which is relevant at all 
temperatures~\cite{kww.80}.

We speculate that the periodic Anderson model 
with a large FK term will exhibit the same behavior 
as the FK model at high temperatures. Indeed, if the conduction band and the 
f-level are gapped, and the width of the f-level is large, then 
the effect of the 
hybridization can be accounted for by renormalizing the parameters
of the FK model. On the other hand, if the low-temperature 
state of the full model is close to the valence-fluctuating fixed point 
with the conduction band and hybridized f-level close to the Fermi 
level, then the likely effect of the FK correlation is to renormalize the 
parameters of the Anderson model. 

\begin{acknowledgments}
We acknowledge discussions with Z. Fisk, B. L\"uthi, 
M. Miljak, M. O\v cko, and J. Sarrao.
This research was supported by the National Science Foundation under grant
DMR-9973225. 
\end{acknowledgments}

\end{document}